\documentclass[
    ,final            
  ]
  {aipproc}

\layoutstyle{8x11double}

\newcommand{\gtrsim}{\mathrel{\hbox{\rlap{\lower.55ex \hbox {$\sim$}}
                   \kern-.3em \raise.4ex \hbox{$>$}}}}
\newcommand{\lesssim}{\mathrel{\hbox{\rlap{\lower.55ex \hbox {$\sim$}}
                   \kern-.3em \raise.4ex \hbox{$<$}}}}

\begin{document}

\title{X-ray spectral and timing properties of the 2001 superburst of 4U\,1636$-$536}

\author{Erik Kuulkers}{
  address={INTEGRAL Science Operations Centre,
       Science Operations and Data Systems Division,
       Research and Scientific
       Support Department of ESA, ESTEC,
       Postbus 299, NL-2200 AG Noordwijk, The Netherlands}
}

\author{Jean in 't Zand}{
  address={SRON National Institute for Space Research, Sorbonnelaan 2, 3584 CA Utrecht, The Netherlands \&\
Utrecht University, PO Box 80000, 3508 TA Utrecht, The Netherlands}
}

\author{Jeroen Homan}{ 
 address={Center for Space Research, Massachusetts Institute of Technology, 77 Massachusetts Avenue, Cambridge, MA 02139, USA}
}

\author{Steve van Straaten}{
 address={Astronomical Institute ``Anton Pannekoek" University of Amsterdam and Center for High-Energy Astrophysics, Kruislaan 403, 1098 SJ Amsterdam, The Netherlands}
}

\author{Diego Altamirano}{
 address={Astronomical Institute ``Anton Pannekoek" University of Amsterdam and Center for High-Energy Astrophysics, Kruislaan 403, 1098 SJ Amsterdam, The Netherlands}
}

\author{Michiel van der Klis}{
 address={Astronomical Institute ``Anton Pannekoek" University of Amsterdam and Center for High-Energy Astrophysics, Kruislaan 403, 1098 SJ Amsterdam, The Netherlands}
}

\begin{abstract}
Preliminary results are reported on the spectral and timing properties of the spectacular 2001 superburst of 
4U\,1636$-$536 as seen by the RXTE/PCA. 
The (broad-band) power-spectral and hardness properties during the superburst
are compared to those just before and after the superburst.
Not all of the superburst emission can be fitted by pure black-body radiation.
We also gathered BeppoSAX/WFC and RXTE/ASM data, as well as other RXTE/PCA data, obtained days to months before 
and after the superburst to investigate the normal X-ray burst behavior around the time of the superburst.
The first normal X-ray burst after the 2001 superburst was detected $\simeq$23~days later.
During inspection of all the RXTE/ASM data we found a third superburst. This superburst took place on June 26, 1999,
which is $\simeq$2.9~yrs after the 1996 superburst and $\simeq$1.75~yrs before the 2001 superburst.
The above findings are the strongest constraints observed so far on the duration of the cessation of normal 
X-ray bursts after a superburst and the superburst recurrence times.
\end{abstract}

\maketitle

\section{Superbursts}

Superbursts are a recently discovered new mode of unstable thermo-nuclear burning
on the surface of a neutron star. These neutron stars reside in so-called
low-mass X-ray binaries (LMXBs) in which it continuously receives fresh material from
a donor star. A large fraction of these LMXBs are known to exhibit short
(seconds to minutes) so-called Type~I X-ray bursts which are due to unstable thermo-nuclear burning
of H and/or He (for reviews, see Lewin et al.\ \cite{Letal1993}; Strohmayer \&\ Bildsten \cite{SB2003}).
These `normal' X-ray bursts appear as rapid ($\sim$1--10\,s) increases in the X-ray flux, followed by
an exponential-like decline.
They recur with a frequency (typically hours to days) which is
(partly) set by the supply rate of the fresh fuel.
The (net) burst X-ray spectra are well described by black-body
emission from a compact object with $\sim$10\,km radius and
color temperature of $\sim$1--2\,keV. The color temperature
increases during the burst rise and decreases
during the decay, reflecting the heating and subsequent cooling of the neutron star surface.
Typical integrated burst energies are in the 10$^{39}$ to 10$^{40}$\,erg range.

The durations and output energies of superbursts are roughly a factor of a thousand 
more than that of the normal X-ray bursts. They display, however, the usual
relatively fast rise and exponential decay, as well as the hardening
during the rise and subsequent softening during the decay, as seen in 
normal X-ray bursts. Also, the (net) superburst X-ray spectra are described
in the same way as the normal X-ray bursts. For an overview of the superburst
properties we refer to Kuulkers \cite{K2003}, and references therein.

The current view is that the superbursts are due to the energy release of
unstable burning of C, which is the left over from the stable and unstable 
burning of H and He (Cumming \&\ Bildsten \cite{CB2001}; Strohmayer \&\ Brown \cite{SB2002}).
Additional energy may be released through photo-desintegration triggered
by the superburst (Schatz et al.\ \cite{Setal2003}).

The first superburst was discovered with the BeppoSAX/WFCs to come from the X-ray 
burster 4U\,1735$-$444 (Cornelisse et al.\ \cite{Cetal2000}). Another superburst was independently found with the
RXTE/PCA, originating from the X-ray burster 4U\,1820$-$303 (Strohmayer \&\ Brown \cite{SB2002}; see also
Ballantyne \&\ Strohmayer \cite{BS2004}).
Thereafter, six more events have been seen to occur in five other X-ray bursters
with the BeppoSAX/WFCs (Cornelisse et al.\ \cite{Cetal2002}; Kuulkers et al.\ \cite{Ketal2002},
see also in 't Zand et al.\ \cite{Zetal2004}), 
RXTE/PCA (Strohmayer \&\ Markwardt \cite{SM2002}) and 
RXTE/ASM (Wijnands \cite{W2001}; Kuulkers \cite{K2002}). 
The fact that only eight such events have been found so far indicates that they
must be rare. Two superbursts from 4U\,1636$-$536 were seen to occur 
$\simeq$4.7~yrs apart (Wijnands \cite{W2001});
observational estimates of the recurrence times are on the
order of a year (see, e.g., Kuulkers \cite{K2002}; in 't Zand et al.\ \cite{Zetal2003}).

4U\,1636$-$536 is a regular X-ray burster
and a so-called `atoll' source (e.g., van der Klis et al.\ 1990).
We here present preliminary results focussed on the X-ray behavior of the source
during and around the superburst which occurred on 2001 Feb 22.

\subsection{The 2001 superburst as seen by the RXTE/PCA}

A first account of the RXTE/PCA data during the 2001 superburst from 4U\,1636$-$536 
is given by Strohmayer \&\ Markwardt \cite{SM2002}. Note that this is event was also seen
by the RXTE/ASM (Wijnands \cite{W2001}; see also below). Strohmayer \&\ Markwardt  \cite{SM2002}
found a highly coherent 
582\,Hz pulsation during a small part ($\sim$800\,s) of the superburst. 

\begin{figure}
  \includegraphics[bbllx=33pt,bblly=70pt,bburx=568pt,bbury=550pt,angle=-90,width=0.48\textwidth]{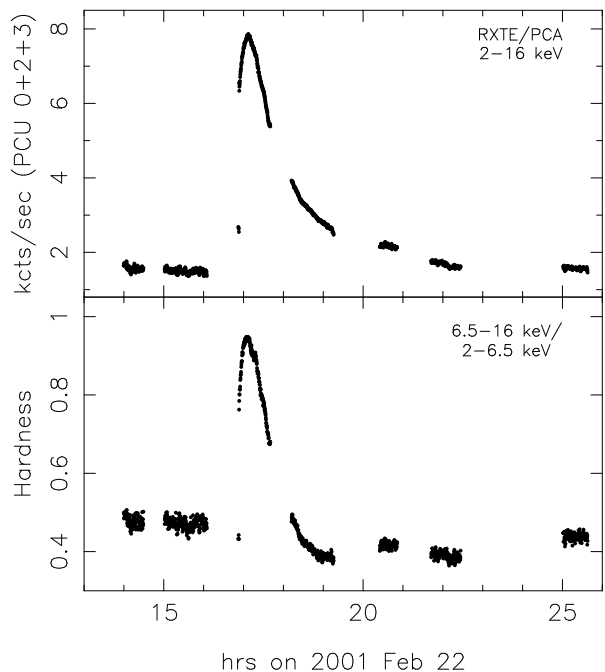}
  \caption{RXTE/PCA 2--16\,keV light curve ({\it Top}) and
hardness (ratio of the count rates in the 6.5--16\,keV and 2--6.5\,keV energy bands) curve
({\it Bottom}) of the 2001 superburst of 4U\,1636$-$536. 
Note that slew data are not included, and that the count rates are not corrected for background and dead time.
Standard 2 data are used from PCU's 0, 2 and 3; the time resolution is 16\,s.}
\end{figure}

The light curve (see also Strohmayer \&\ Markwardt \cite{SM2002}), as well as the hardness curve, of the superburst are 
shown in Fig.~1. Clearly, they resemble the curves usually seen for normal X-ray bursts, except 
for the duration of a few hours. A high time resolution (0.125\,s) light curve of the 
superburst is shown by Kuulkers \cite{K2003}, including data from the slew just before the third orbit of data
on 4U\,1536$-$536. This light curve showed an increase in count rate of a few 1000\,c\,s$^{-1}$ about
125\,s before the precursor X-ray burst. It was pointed out to us by Tod Strohmayer that this 
increase is actually due to the turn-on of PCU 3; this was not taken into account when doing the
collimator response correction. The superburst, therefore, did {\it not} start $\sim$125\,s before
the precursor X-ray burst, but between $\sim$145\,s and $\sim$46\,min before the precursor.

\begin{figure}
  \includegraphics[bbllx=78pt,bblly=23pt,bburx=587pt,bbury=702pt,angle=-90,width=0.48\textwidth]{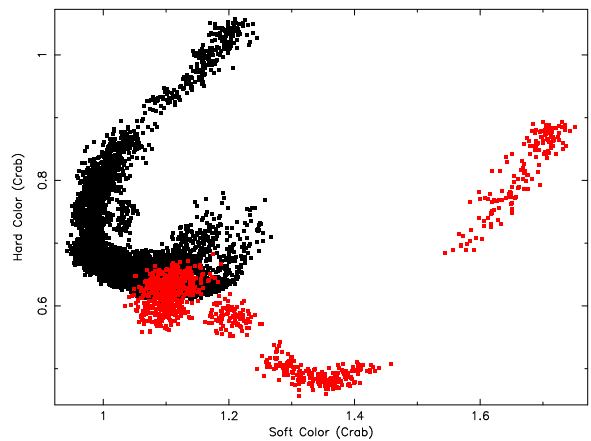}
  \caption{Color-color diagram of 4U\,1636$-$536 using Standard 2 data between Feb 28, 1996 and April 10, 2002.
The data on Feb 22, 2002 are indicated in red. Soft and hard color are defined as the ratio of the count rates 
in the 3.5--6.4\,keV and 2.0--3.5\,keV bands, and, respectively, the 9.7--16\,keV and 6.4--9.7\,keV bands.
The count rates are normalized to the Crab (see Di Salvo et al.\ \cite{DSetal2002}; van Straaten et al.\ \cite{Setal2003}) 
and not corrected for dead time.}
\end{figure}

A color-color diagram (CD) of all public data on 4U\,1636$-$536 between Feb 28, 1996 and April 10, 2002
is shown in Fig.~2 (this is an extension of the data set used by Di Salvo et al.\ \cite{DSetal2002}).
The superburst (shown in red) started at the bottom part of the CD, i.e., in the
so-called `banana' branch. During the rise to
superburst peak the source hardened, while during the decay it softened, with the hard colors 
even dropping below pre-superburst values. At the end of the observation the source resumed its
pre-superburst colors again.

We investigated the broad-band timing behavior using the high-time resolution modes available 
during parts of the observation (see Strohmayer \&\ Markwardt \cite{SM2002}). No strong features are seen
before and after the superburst, except for some very-low frequency noise. This is as expected
from previous analyses of the timing behavior at the corresponding portion of the CD
(see Di Salvo et al.\ \cite{DSetal2002}). The high-timing data during the superburst, however, show 
a hint of a quasi-periodic oscillation near 1150\,Hz. A more careful analysis is in progress.

Using the Standard 2 mode data we extracted X-ray spectra at 16\,s time resolution throughout the
2001 superburst. As is common practice for normal X-ray bursts, we subtracted the average pre-superburst
X-ray spectrum from the superburst spectra. The residual spectra were modeled by pure black-body emission
(keeping the interstellar absorption, $N_{\rm H}$, fixed at 3$\times$10$^{21}$~atoms\,cm$^{-2}$, see e.g.,
Christian \&\ Swank \cite{CS1997}).
As the superburst flux increased the color temperature increased from $\simeq$1\,keV up to
2.35\,keV, while during the decay the color temperature decreased again.
The inferred black-body radius was more or less constant ($\simeq$6\,km at 6\,kpc).
The fits are, however, not perfect throughout the whole burst, with the strongest deviations 
seen when the superburst flux had dropped to about half its peak value. The deviations
are very similar to those seen during the superburst from 4U\,1820$-$303, and may show the
influence of the superburst emission on the accretion disk surrounding the neutron star
(Strohmayer \&\ Brown \cite{SB2002}; Ballantyne \&\ Strohmayer \cite{BS2004}).
The total integrated superburst flux is about 0.65$\times$10$^{42}$\,erg (at 6\,kpc).

\subsection{BeppoSAX/WFC and RXTE/ASM measurements around the 2001 superburst}

\begin{figure}
  \includegraphics[bbllx=52pt,bblly=50pt,bburx=553pt,bbury=552pt,angle=-90,width=0.48\textwidth]{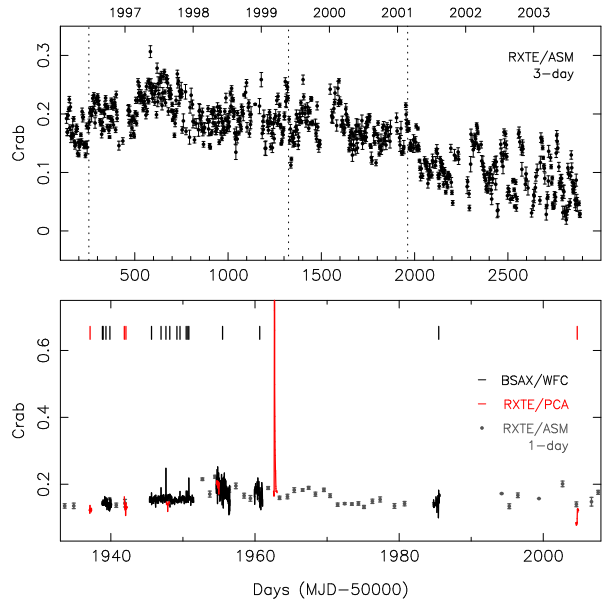}
  \caption{{\it Top}: RXTE/ASM (1.5--12\,keV) light curve of 4U\,1636$-$536 from Jan 6, 1996 to 
Sep 4, 2003. The data shown are 3-day averages and normalized to the Crab count rate.
With dotted lines we indicate the times of the 3 superbursts seen from 4U\,1636$-$536 (see text).
{\it Bottom}: The BeppoSAX/WFC (2--28\,keV) light curve at 15-min time resolution during the interval Jan 21 to 
Apr 9, 2001 is shown in black. Data which were less than 0.5~day apart were connected to guide the
eye. They were normalized to the Crab count rate. In gray we show the RXTE/ASM 1-day average data points, 
as well as in red the RXTE/PCA 15-min average light curves, in the same period; within an RXTE/PCA observation
the data are connected. In the top part of the bottom panel we show with vertical bars the times of occurrence of 
normal X-ray bursts.}
\end{figure}

In Fig.~3 (bottom) we show in black the WFC light curves acquired around the 2001 superburst of
4U\,1636$-$536, as well as in gray the RXTE/ASM 1-day average and in red the RXTE/PCA light curves.
The BeppoSAX/WFC observations did not cover the superburst; the RXTE/ASM, however, did cover it
(Wijnands \cite{W2001}; see also below). In the top of the bottom panel figure we annotate with bars the
times of normal X-ray bursts as seen by the BeppoSAX/WFC (black) and the RXTE/PCA (red).
Clearly, the source showed normal X-ray bursts before the superburst, with an average rate of about 
6 per day. Observations after the 2001 superburst are sparse; the first BeppoSAX/WFC and RXTE/PCA observations
were $\sim$23~days, respectively, $\sim$42~days after the superburst. During both observations 
a normal X-ray bursts was seen. We note that the peak count rates during the normal X-ray bursts
are a factor of $\sim$1.5--2 higher than that reached during the precursor burst during the superburst.
This is consistent with an early ignition of a H/He layer by the superburst
as it starts (Cumming \cite{C2003}).

\subsection{RXTE/ASM superbursts}
\label{asm}

In the top panel of Fig.~3 we show the long-term RXTE/ASM 3-day average light curve of 4U\,1636$-$536.
The source varies on time scales of days to months to years. It may be interesting to note that
this light curve resembles that of KS\,1731$-$260 during the first 4 year of the RXTE mission.
Apart from the 2001 event, a superburst was also seen by the RXTE/ASM on June 19, 1996 (Wijnands \cite{W2001}).
In Fig.~4 we show the latter event on the left, whereas the 2001 event is shown on the right.
During the inspection of the individual RXTE/ASM dwell light curves (see, e.g., Wijnands \cite{W2001}),
we found, however, another candidate superburst. This candidate, which occurred on June 26, 1999, 
is shown in the middle panels of Fig.~4. Although the peak of the light curve (top) consists of
1 data point only, the data points after that are on average higher than data before and after.
To guide the eye, we have connected the data points; the light curve of the candidate
is consistent with a fast rise plus slower exponential decay light curve.
In the bottom panel we show the hardness curve in the same period; the hardness values
during the peak and during the decay, as well as their evolution, are very similar to those of the superbursts
in 1996 and 2001. Based on the similarities we regard the 1999 event as a real superburst; if true,
then the recurrence times between the 1996 and 1999 events and the 1999 and 2001 are $\simeq$2.9~yrs and
$\simeq$1.75~yrs, respectively. 

\begin{figure}
  \includegraphics[bbllx=33pt,bblly=70pt,bburx=573pt,bbury=770pt,angle=-90,width=0.48\textwidth]{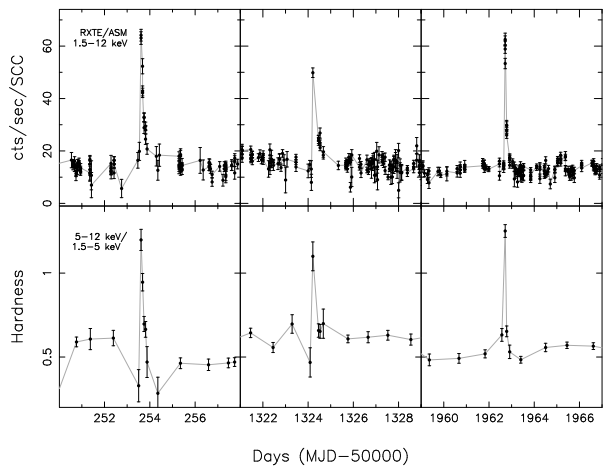}
  \caption{RXTE/ASM 1.5--12\,keV light curves ({\it top}) and hardness curves ({\it bottom}) around the
superbursts seen from 4U\,1636$-$536 in 1996 (Wijnands \cite{W2001}; {\it left}), 1999 (see text; {\it middle}) and 2001
(Wijnands \cite{W2001}; Strohmayer \&\ Markwardt \cite{SM2002}; {\it right}). 
For the light curves we show the individual 90-s average 
RXTE/ASM dwell data points. Hardness is defined as the ratio of the count rates in the 
5--12\,keV and 1.5--5\,keV bands; values are averages of nearby dwells. Data points are connected in
light gray to guide the eye.}
\end{figure}

\section{Summary}

In this proceedings paper we have shown some preliminary results on the analysis of 
RXTE/PCA, RXTE/ASM and BeppoSAX/WFC data obtained during and around the 2001 superburst
seen from the X-ray burster 4U\,1636$-$536. We found that during the first observation
after this superburst, i.e., 23 days later, a normal X-ray burst occurred. Previously, we found that
normal X-ray bursts seem to cease to occur for up to a month after the superbursts of
Ser\,X-1 (Cornelisse et al.\ \cite{Cetal2002}) and KS\,1731$-$260 (Kuulkers et al.\ \cite{Ketal2002}).
Our observed upper limit on the duration of the cessation of normal X-ray bursts after the 2001 superburst 
puts strong constraints on the modeling of the thickness of the layer where the 
superburst originates as well as where the energy is deposited
(Cumming \&\ Macbeth \cite{CM2004}).

We also investigated the entire RXTE/ASM data set on
4U\,1636$-$536 and found evidence for a third superburst, which occurred in 1999
(the first superburst occurred in 1996). The shortest observed superburst recurrence time
is now $\simeq$1.75~yrs, which is more or less in agreement with previous estimates
(roughly one superburst per source per year; Kuulkers \cite{K2002}; in 't Zand et al.\ \cite{Zetal2003}).
Superburst recurrence times may constrain the original composition of the material
which was deposited on the neutron star (Cumming \cite{C2003}; see also 
Cumming \cite{C2003b}, and the discussion in Kuulkers \cite{K2003}).

\begin{theacknowledgments}
  We thank Tod Strohmayer for pointing out to us about the turn-on of the 
various PCU's during the slew towards the source when the superburst occurred.
\end{theacknowledgments}

\bibliographystyle{aipproc}   


\end{document}